\documentclass[twocolumn,amsmath,amssymb,floatfix,prb,shownopacs,footinbib,superscriptaddress]{revtex4-1}
\usepackage{amsmath,amssymb,natbib,bm,graphicx,url,epsf,color}
\usepackage[british]{babel}
\usepackage{wasysym}
\usepackage{MnSymbol}

\begin{document}

\title{Direct probing of band-structure Berry phase in diluted magnetic semiconductors}

\author{M. Granada}
\altaffiliation[Permanent address: ]{Centro At\'omico Bariloche, CNEA, (8400) San Carlos de Bariloche, Argentina.}
\affiliation{CNRS, Laboratoire de Photonique et de Nanostructures, route de Nozay, 91460 Marcoussis, France.}
\affiliation{Consejo Nacional de Investigaciones Cient\'ificas y T\'ecnicas, Argentina.}

\author{D. Lucot}
\affiliation{CNRS, Laboratoire de Photonique et de Nanostructures, route de Nozay, 91460 Marcoussis, France.}

\author{R. Giraud}
\affiliation{CNRS, Laboratoire de Photonique et de Nanostructures, route de Nozay, 91460 Marcoussis, France.}

\author{A. Lema\^{\i}tre}
\affiliation{CNRS, Laboratoire de Photonique et de Nanostructures, route de Nozay, 91460 Marcoussis, France.}

\author{C. Ulysse}
\affiliation{CNRS, Laboratoire de Photonique et de Nanostructures, route de Nozay, 91460 Marcoussis, France.}

\author{X. Waintal}
\affiliation{SPSMS, UMR-E 9001, CEA-INAC/ UJF-Grenoble 1,17 rue des Martyrs, 38054 Grenoble cedex 9, France.}

\author{G. Faini}
\email[Author to whom correspondence should be addressed: ]{giancarlo.faini@lpn.cnrs.fr}
\affiliation{CNRS, Laboratoire de Photonique et de Nanostructures, route de Nozay, 91460 Marcoussis, France.}

\date{\today}

\begin{abstract}

We report on experimental evidence of the Berry phase accumulated by the charge carrier wave function in single-domain nanowires made from a (Ga,Mn)(As,P) diluted ferromagnetic semiconductor layer. Its signature on the mesoscopic transport measurements is revealed as unusual patterns in the magnetoconductance, that are clearly distinguished from the universal conductance fluctuations. We show that these patterns appear in a magnetic field region where the magnetization rotates coherently and are related to a change in the band-structure Berry phase as the  magnetization direction changes. They should be thus considered as a band structure Berry phase fingerprint of the effective magnetic monopoles in the momentum space. We argue that this is an efficient method to vary the band structure in a controlled way and to probe it directly. Hence, (Ga,Mn)As appears to be a very interesting test bench for new concepts based on this geometrical phase.

\end{abstract}

\maketitle

\doi{10.1103/PhysRevB.91.235203}

\section{INTRODUCTION}

A quantum system undergoing an adiabatic evolution accumulates a geometrical phase, which adds to the standard dynamical phase. This  phase, introduced by Berry in his seminal paper \cite{Berry_PRL84}, has become a central concept in our understanding of quantum mechanics. The signature of this Berry phase has been sought in a broad range of domains in physics, like in neutron beams \cite{Bitter_PRL87}, NMR measurements in molecules\cite{Jones_Nature2000} and solid-state superconducting qubits \cite{Leek_Science07}. In condensed matter, the Berry phase is intimately related to band theory. It provides a deep understanding of, say, the topological nature of the quantum Hall effect or some peculiar features of graphene \cite{Zhang_Nature05, Liu_PRL11}. Despite its fundamental nature, no direct experimental signatures have been reported, except through reinterpretation of anomalous Hall-related effects \cite{Jungwirth02, Fang_Science03, Onoda_PRB08,  Glunk_PRB09, Werpachowska_PRB10}.

One of the cornerstones of condensed-matter physics is the Bloch theorem and the associated concept of band structure.
Bloch's theorem states that, inside a crystal, the electronic wave function with momentum $\vec k$ in a band $n$ takes the form
$\Psi (\vec r)= e^{i\vec k.\vec r} u_{n\vec k}(\vec r)$
where the plane wave part $e^{i\vec k.\vec r}$ describes the long distance physics of the system while
$u_{n\vec k}(\vec r)$ captures the short (atomic) distance physics. When $\vec k$ varies (due to elastic collisions with defects, in the present study), the wave function picks up the so called \textit{band structure Berry phase} \cite{Berry_PRL84},
$\gamma=i\int d\vec k.\langle u_{n\vec k}|\nabla_{\vec k} |  u_{n\vec k}\rangle$.
This Berry phase is therefore ubiquitous in condensed matter. However, up to now, it has been revealed only indirectly. Two ingredients are indeed required to observe it: a quantum interferometer to probe any phase changes and an adjustable parameter to control the Berry phase accumulated by the carrier wave function upon traversing closed paths on the Fermi surface. 
Universal conductance fluctuations (UCFs) in disordered materials are a well-known manifestation of quantum interference effects. These aperiodic but reproducible UCFs appear as an external magnetic field is varied \cite{Lee_PRB87} and have been extensively used as a probe of the phase coherence in mesoscopic samples since the mid-'80s. The fluctuation pattern is a signature of quantum interference between different electronic trajectories in real space, reflecting the singular microscopic configuration of disorder for a given sample, yielding the so-called ``magneto-fingerprint" of that individual sample. The mean value of the amplitude of these fluctuations is universal, and of the order of the quantum of conductance, $e^2/h$. Hence, UCF experiments can be viewed as a ``poor man's quantum interferometer" picking up the long distance particular physics of the system. We need now a tunable parameter to modify the Berry phase. As mentioned above, this phase is intrinsically related to the band structure, so the accumulated Berry phase is related to the band configuration at the Fermi surface. In (Ga,Mn)As compounds the valence band is split due to the effective field arising from the exchange interaction between the hole spin and the polarized Mn moments.\cite{bands} Because of the spin-orbit coupling, the band splitting is highly anisotropic with respect to the direction of magnetization. Therefore the magnetization, and more specifically its orientation, appears to be a relevant parameter to produce changes in the band structure, and hence in the associated Berry phase acquired by the quasi-particle wave function.

Phase-coherent transport has been investigated in (Ga,Mn)As devices over the past few years:  UCF\cite{Wagner_PRL06, Vila_PRL07} and non-locality effects\cite{Giraud_ASS07} were reported. Also experimental evidence for weak localization\cite{Neumaier_PRL07, Neumaier_NJPhys2008}, as well as theoretical studies\cite{Garate_PRB09}, has been presented. It was found that low energy spin wave modes hardly affect the phase coherence. As shown in Ref. [\onlinecite{Vila_PRL07}], neither magnetization rotation nor magnetic texture destroy coherence in those systems. More strikingly, very unusual patterns, consisting of fast fluctuations of the conductance, were observed at low fields in addition to the conventional UCFs\cite{Vila_PRL07}. This unexpected regime appeared as the magnetic field was swept perpendicular to the easy axis of magnetization, below the anisotropy field. 
Such behavior was conjectured to be related to the presence of magnetic domain walls (DWs), yielding to an additional phase term connected to the magnetic disorder. However, in those experiments, the knowledge of the sample magnetic configuration was not accessible during the transport measurements, and no conclusive argument could be provided. More recently, Hals and co-workers\cite{Hals_PRL10} proposed a new physical mechanism accounting for the existence of this regime. The fast oscillations would be the Berry phase fingerprint of effective magnetic monopoles in momentum space, which arise from energy-band crossings\cite{BohmSPRING03}. When a quasi-particle wave function traverses a closed loop in momentum space, it accumulates a geometrical phase from the monopole field.  In (Ga,Mn)As the position of these energy-crossings, or monopoles, depends on the magnetization orientation, due to the strong spin-orbit coupling. Hence, magnetization reorientation leads to a relocation of the monopoles in $k$-space, which in turn yields a Berry phase change.

These conflicting interpretations called for further investigations. To that end, we designed a new sample geometry so that the magnetic state could be well established. UCF experiments were performed for different magnetic field orientations and temperatures. As we will show later, our results indicate that the DW contribution can be clearly discarded as being the origin of these fast fluctuations, and point toward a fingerprint of the band structure Berry phase.
\section{DESCRIPTION OF THE SAMPLES AND EXPERIMENTAL SETUPS}

\begin{figure}[!htbp]
\centering\includegraphics[width=1.0\columnwidth]{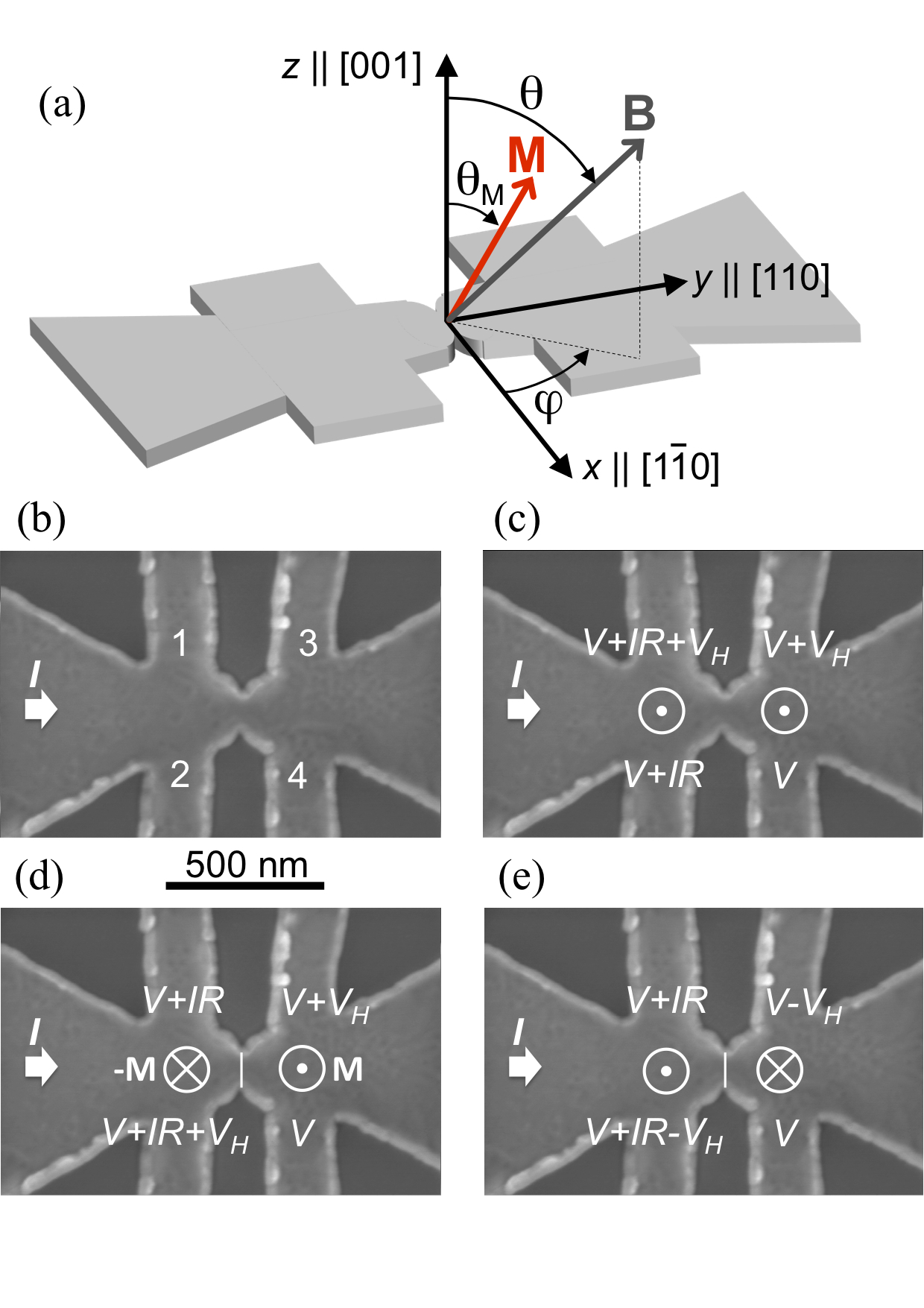}
\caption{(Color online.) (a) Schematic view of the orientation of crystallographic axes of the sample and applied field direction angles $\theta$, $\varphi$. \textbf{B} denotes the applied magnetic field, \textbf{M} denotes magnetization and $\theta_{M}$ the magnetization angle from the easy axis [001]. In the present work, $\varphi$ is fixed to $90^\circ$. (b) Scanning electron microscope image of a Hall bar indicating the numbering of voltage contacts and the direction of electric current $I$. (c)-(e) The voltage at each contact is indicated for different magnetization configurations. $V_H$ is the Hall voltage, $IR$ is the longitudinal voltage drop and $V$ is an offset voltage referred to ground. (c) The magnetization is homogeneous throughout the sample. (d)-(e) There is a DW in the constriction, with the magnetization presenting opposite sign at both sides.
 } 
\label{Fig.1}
\end{figure}

Our experiments were performed on a ferromagnetic semiconductor thin film. A 50 nm thick (Ga$_{0.95}$Mn$_{0.05}$)(As$_{0.89}$P$_{0.11}$) epilayer was grown at 250 $^\circ$C by molecular beam epitaxy on a (001) GaAs substrate. 
Doping (GaMn)As with phosphorus produces a variation of the cell parameter of the compound and, in turn, the substrate induced strains are modified. This allows us to tune the magnetic anisotropy of (GaMn)(AsP) epitaxial layers deposited directly on GaAs substrates by changing only the phosphorus concentration\cite{Lemaitre_APL08}. The film used in this work presents an out-of-plane easy axis of magnetization.
After conventional annealing treatment, the Curie temperature was $T_C=$ 113 K and the sheet resistance $R_\square\sim1.7$ k$\Omega$. This film exhibits a low density of DW pinning centers.\cite{Haghgoo_PRB2010} Submicrometric Hall bars were defined on the wafers using electron beam lithography and ion beam etching, with the geometry shown in Fig. \ref{Fig.1}. 
The distance between the voltage contacts for longitudinal conductance measurements is $L=$ 440 nm, comparable to the phase coherence length measured at low temperature for similar compounds in previous works ($L_\Phi \sim 100$ nm at $T=100$ mK)\cite{Vila_PRL07}. An $\sim$100 nm wide constriction in the middle of the bar acts as an efficient pinning center for DWs; this constriction was designed to trap domain walls at will and to probe their influence on the dynamical phase. Finally, Ti/Au (20/200nm) Ohmic contacts were thermally evaporated.

\subsection{Domain wall detection}

\begin{figure*}[!htbp]
\centering\includegraphics[width=0.75\textwidth]{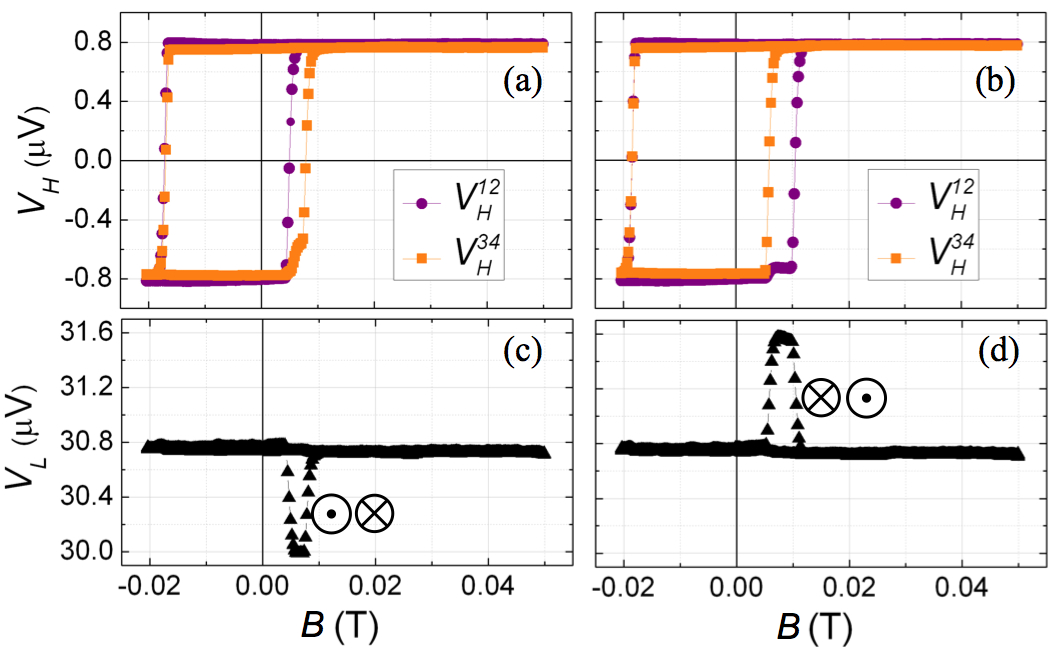}
\caption{(Color online.) (a) Hall voltages measured on the left (circles) and right (squares) of the constriction throughout a field sweep at $T=4$ K. In the region between the different positive coercive fields, the Hall voltages on both sides of the constriction with opposite sign indicate that a domain wall is present in the constriction. (b) Another realization of the same experiment displaying the opposite order in the magnetization reversal on both sides of the constriction. (c)-(d) Longitudinal voltage, $V_L^{24}$, measured simultaneously with the Hall voltages displayed on top. The presence of a domain wall can be detected in the longitudinal voltage due to a Hall effect contribution ($IR\pm V_H$ in Fig.\ref{Fig.1}(d) and (e)). The left panels ((a) and (c)) present a measurement with the magnetization configuration described in Fig. \ref{Fig.1}(e), while the results displayed in the right panels ((b) and (d)) correspond to the case in Fig. \ref{Fig.1}(d). 
 } 
\label{Fig.2}
\end{figure*}

The double Hall bar geometry (Fig. \ref{Fig.1}) allows for detection of a domain wall (DW) trapped in the constriction by comparing the Hall voltages on both sides of the constriction. Indeed, in (Ga,Mn)(As,P), the Hall voltage ($V_H$) is completely dominated by the anomalous Hall effect, which is proportional to the local magnetization component perpendicular to the sample plane. Therefore, $V_H$($B$) reproduces the magnetization loops. The presence of a DW at or close to the constriction would result in Hall voltages with opposite sign on both sides of the constriction.

If there are no domain walls, meaning that the magnetization is homogeneous throughout all the sample, the Hall voltage on the left should be the same as that on the right, that is, $V_H^{1,2} = V_H^{3,4}$ (see Fig. \ref{Fig.1}(b) for the labeling of electrical contacts).
In this case, the longitudinal voltage will be $V_L^{2,4} = V_L^{1,3} = I.R$, with $R$ being the resistance between contacts $2$ and $4$ (see Fig. \ref{Fig.1}(c)). On the contrary, if there is a DW trapped in the constriction, the magnetization will have opposite sign at both sides. In this case, $V_H^{1,2} = - V_H^{3,4}$, and the longitudinal voltage will be $V_L^{2,4} = I.R \pm V_H$, with the plus or minus sign depending on the magnetization configuration corresponding to Fig. \ref{Fig.1}(d) or (e).

Figure \ref{Fig.2} shows the effect of a DW pinned in the constriction, on both the Hall (Fig. \ref{Fig.2}(a) and \ref{Fig.2}(b)) and on the longitudinal (Fig. \ref{Fig.2}(c) and \ref{Fig.2}(d)) measured voltages. When the magnetic field is swept from the positive to the negative saturation value, simultaneous magnetization reversal occurs on both sides of the constriction, yielding to the reversal of both $V_H$ curves for the same negative coercive field. Nucleation and propagation of domain walls are the well-established reversal mechanism in these materials\cite{Gould_PRL04}, so DWs are present in this process. However, since the nucleation energy is higher than the depinning one, no DWs are trapped in the constriction. Thus, once DWs are created from a saturated state, they have enough energy to travel across the constriction. If the magnetic field is swept to high negative values, the magnetization will be saturated and the same effect will be observed when sweeping back to positive fields, with the reversal on both sides of the constriction occurring at the same positive field value. However, a different effect occurs if the field is not swept beyond $-20$ mT, as shown in Fig. \ref{Fig.2}. In this case, the magnetization is not saturated and DWs must still be present in the sample, although far away from the voltage contacts. Then, when sweeping the field back to positive values, the first coercive field is observed when one of the $V_H$ signals changes sign ($V_H^{1,2}$ and $V_H^{3,4}$ in Figs. \ref{Fig.2}(a) and \ref{Fig.2}(b) respectively), meaning that a DW has propagated and gotten trapped in the constriction. By sweeping further the magnetic field, there is a field range where the DW remains trapped, and then the second $V_H$ curve reverses. The left and right panels in Fig. \ref{Fig.2} correspond to the opposite magnetization configurations depicted in Figs. \ref{Fig.1}(d) and (e). Furthermore, not only does $V_H$ indicates the presence of a DW, but $V_L$ is also sensitive to this effect, as shown in the bottom panels of Fig. \ref{Fig.2}. Comparing the top and bottom panels, the relationship $V_L^{2,4} = I.R \pm V_H$ can be deduced, where the minus sign corresponds to Figs. \ref{Fig.2}(a) and (c), and the plus sign corresponds to Figs.\ref{Fig.2}(b) and (d).
We have proved here that, despite the lack of control on the magnetic configuration (that is, we cannot choose which configuration to stabilize), we can pin and detect a DW, and also distinguish between the two possible configurations.
 Similar experiments were performed in the mesoscopic regime, i.e., at temperatures lower than 1 K. In this case, the magnetization reversal is clearly observed in $V_H$, although conductance fluctuations may veil the associated jumps in $V_L$. 

After validating the detection procedure of DWs, this method was used to study the magnetization configuration with the magnetic field applied at different angles $\theta$ from the easy axis. A thorough analysis of the magnetization reversal processes in this Hall bar was carried out prior to the UCF experiments: we observed that after saturating the magnetization at B = 1 T, the field can be swept back and forth without any DW being created, as long as the field is always kept positive. This allowed us to work in a single magnetic domain configuration while applying the magnetic field at any angle $\theta < 90^{\circ}$, which was important in order to get rid of any domain wall contribution to the magnetotransport experiments.

\subsection{Conductance in the mesoscopic regime}

\begin{figure}[!hbtp]
\centering\includegraphics[width=1.0\columnwidth]{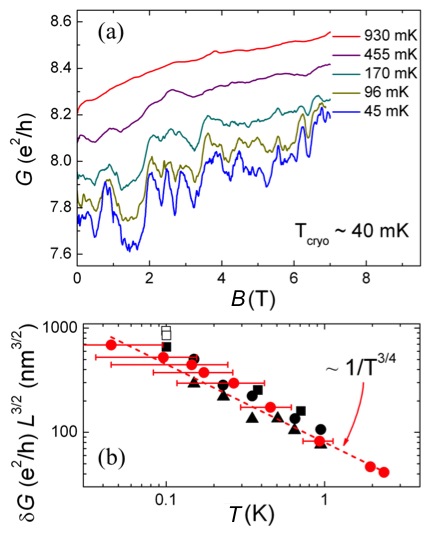}
\caption{(Color online.) (a) Magnetoconductance measured in a 4-probe configuration at different temperatures, with the magnetic field applied along the magnetization easy axis. (b) Temperature dependence of the rms amplitude $\delta G$ normalized to the distance between contacts, \textit{L}. Red symbols are data obtained in this work for GaMnAsP, and black symbols are data presented in Ref. [\onlinecite{Vila_PRL07}], different symbol shapes corresponding to samples of different sizes.
 } 
\label{Fig.3}
\end{figure}

Electronic transport experiments were performed in a dilution refrigerator with a base temperature of 40 mK using a standard four-probe lock-in technique. Variable magnetic field was provided by a 3-axis superconducting magnet system, with a high field up to 7 T for the principal axis $z$ and a 1 T vector field using any combination of $x$-, $y$- and $z$-axis coils.
Since mesoscopic transport had never been measured in this phosphated compound, we first verified the appearance of UCFs. In Fig. \ref{Fig.3}(a), magneto-conductance curves measured at different temperatures are presented.\cite{TvsI} The magnetic field was applied perpendicular to the sample's plane, i.e., along the easy axis of magnetization. In this configuration, with the magnetic field applied along the easy axis, the magnetization remains collinear to the field direction, even at low field. Hence, the band structure remains mostly unchanged over the full field excursion, except for a weak contribution of the band Zeeman effect.

The amplitude of the fluctuations decreases with increasing temperature. 
The behavior of $\delta G.L^{3/2}$ vs. $T$ is shown in Fig. \ref{Fig.3}(b). Here, $\delta G$ is quantified as the rms value $\delta G = \sqrt{\left<[G(B)-\left<G\right>]^2\right>}$ and $L$ is the distance between contacts. The obtained values and the $T^{-3/4}$ dependence, as well as the values of the correlation fields, are in coincidence with those obtained previously in similar mesoscopic devices but made on GaMnAs layers\cite{Vila_PRL07} (shown in Fig. \ref{Fig.3}(b)) , meaning that the intrinsic transport properties are roughly the same, despite the higher concentration of impurities in the present sample.

\section{RESULTS}

\begin{figure}[!htbp]
\centering\includegraphics[width=1\columnwidth]{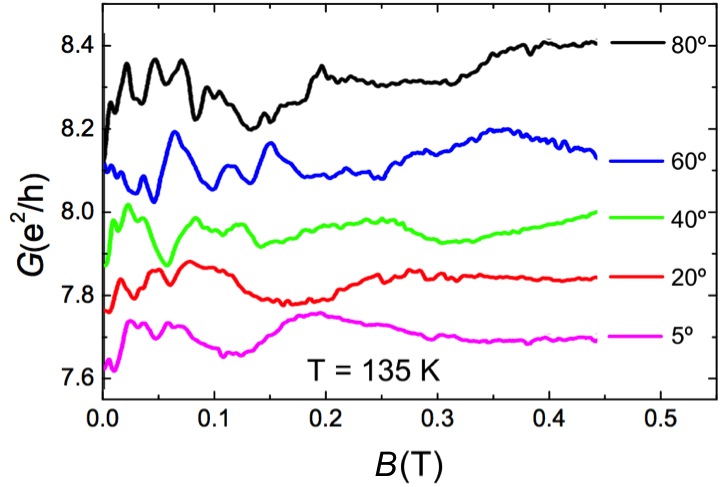}
\caption{(Color online) Magnetoconductance measured at $T = 135$ mK with the magnetic field applied at different angles $\theta$ from the easy axis of magnetization. The curves for $\theta > 10^\circ$ are vertically shifted for clarity.
}
\label{Fig.4}
\end{figure}

To probe the effect of magnetization reorientation on the conductance fluctuations, the magnetic field has to be oriented away from the easy axis. Figure \ref{Fig.4} shows $G(B)$ measured at constant temperature, with the magnetic field applied at different angles $\theta < 90^\circ$. The field was swept between 0 and 1~T.

As mentioned above, in these experiments the magnetization remains in a single domain configuration. Hence, during a field sweep with $\theta > 0^{\circ}$, the magnetization direction ($\theta_{M}$) changes with the magnetic field strength: at low field, the magnetization rotates coherently toward the field direction. After reaching the saturation field, the magnetization remains collinear to the field direction. Above 0.2~ T, where the magnetization is aligned with the field, the conductance exhibits slow fluctuations similar to those in Fig.~\ref{Fig.1}, ascribed to the usual UCF. Below 0.2~T, a new regime develops as the field is rotated away from the easy axis: fast fluctuations are visible. As $\theta$ increases, the fast fluctuations develop within a wider field range. This broadening is consistent with the increasing saturation field, as the magnetic field is tilted away from the easy axis. This first series of experiments clearly confirms that the reorientation of the magnetization produces an additional term in the phase accumulated by the carriers over their trajectories. Hals \textit{et al.}\cite{Hals_PRL10} attributed these fast fluctuations to the signature of a Berry phase, and we will give further evidence supporting this interpretation, by studying the temperature dependence of fluctuations in both regimes.

Let us briefly summarize what is observed in the case of conventional UCFs. As stated before, a UCF fingerprint can be viewed as a ``poor man's interferometer", where the randomness of the fingerprint comes from the lack of control on the different paths that are actually interfering. The relevant length scale is given by the phase coherence length of the charge carriers, $L_\Phi$, which sets the maximum size of the loops that can contribute to the interference pattern. When  $L_\Phi$ is larger than the system size, the UCF amplitude is of the order of $e^2/h$. But as the temperature increases, $L_\Phi$ gets smaller due to decoherence effects, leading to a reduction of the UCF amplitude as shown in Fig. \ref{Fig.3}.  More interesting for us is the typical quasi-period of the $G(B)$ oscillations, which is related to the dynamical phase accumulated in a loop of size $L_\Phi$, which is given by $\Phi= BL_\Phi^2/ h.e$. As the temperature increases, the fluctuations become slower (i.e., the quasi-period increases) since a larger field is needed to put one quantum of flux inside a smaller loop. On the contrary, the band structure Berry phase $\Phi_B$ depends only on the direction of the magnetization and is independent of $L_\Phi$. This is because the relevant trajectory is described in $k$ space where the size of the loop, given by $k_F$, remains nearly unchanged. A simple corollary is that the quasi-period of the oscillations due to a change of the Berry phase is independent of temperature. 

\begin{figure}[!t]
\centering\includegraphics[width=1.0\columnwidth]{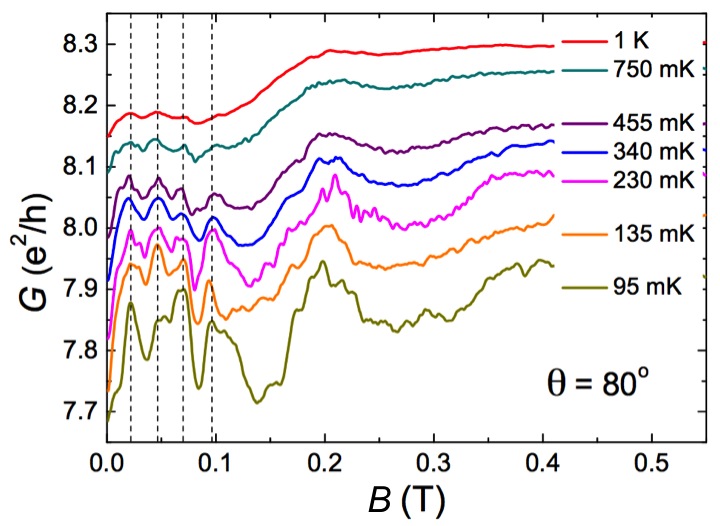}
\caption{ (Color online) Magnetoconductance measured at different temperatures with a fixed magnetic field direction, with $\theta = 80^\circ$. The curves for $T > 95$ mK are vertically shifted for clarity. The vertical gray lines are a guide to the eye, indicating the position of consecutive maxima of $G$.
}
\label{Fig.5}
\end{figure}

To test whether or not these fast fluctuations are a fingerprint of a Berry phase, we performed a series of experiments at different temperatures. Figure \ref{Fig.5} shows $G(B)$ curves measured at different temperatures and fixed magnetic field angle  $\theta = 80^\circ$, for a narrow field window ($0-0.5$ T). Again, both regimes can be clearly identified, for all the temperatures up to 1~K. The fast fluctuations pattern in the low field region, persists as the temperature increases, except for a damping of the amplitude. Moreover, the peaks occur at the same magnetic field values, whatever the temperature. In this regime, the oscillation pseudo-period is therefore temperature independent, contrary to what is observed in the high field UCF regime.

\begin{figure}[!htbp]
\centering\includegraphics[width=1\columnwidth]{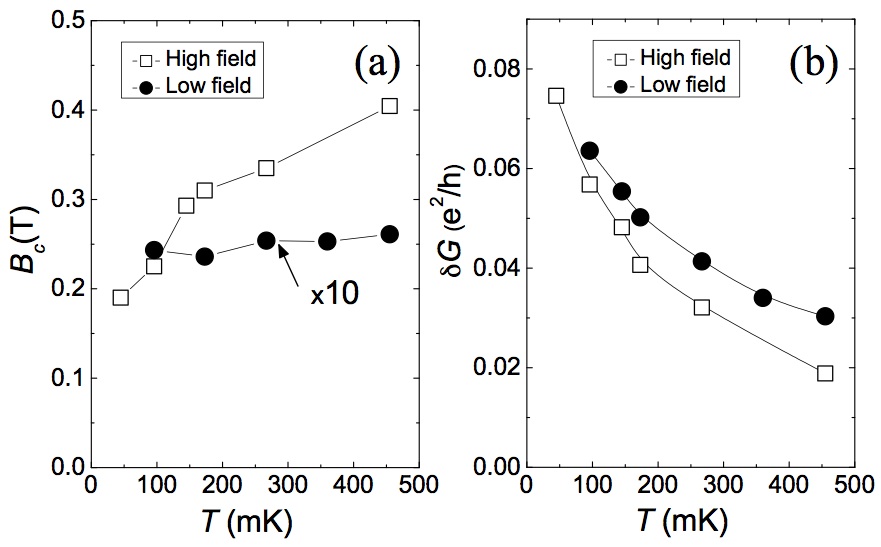}
\caption{Temperature dependence of (a) the pseudo-period $B_c$ and (b) the amplitude $\delta G$ of the fluctuations in the high-field ($\square$) and low-field ($\bullet$) regimes. $B_c$ of the low-field regime is multiplied by 10.}
\label{Fig.6}
\end{figure}

A better insight into the temperature dependence of both regimes is given by the comparison of two relevant parameters: the fluctuation pseudo-period $B_c$ and the rms amplitude of the fluctuations $\delta G$. In the high-field regime, the correlation field  $B_c$ was estimated by calculating the autocorrelation function of $G(B)$ from measurements at $\theta = 0^\circ$ between 0 and 7 T (like those of Fig. \ref{Fig.3}(a)). In the low-field regime, the analysis of the curves in Fig. \ref{Fig.5} ($\theta = 80^\circ$) was restricted to the field range 0-200 mT. In such a limited range, the measured curve displays only a few peaks and valleys and the statistics is not good enough for the autocorrelation function to be calculated; the pseudo-period was thus estimated as the mean distance between consecutive minima in each curve. The latter method was also applied in the high-field regime, in order to check its validity, giving roughly the same temperature dependence as that of the calculated autocorrelation function. The rms amplitude of the conductance fluctuations was calculated as $\delta G = \sqrt{\left<(G(B)-\left<G\right>)^2\right>}$ for both regimes.
The results are summarized in Figure \ref{Fig.6}. 

\section{Discussion}

The evolution of $B_c$ with temperature (Fig.~\ref{Fig.6}(a)) shows clear evidence that the two oscillating regimes have different origins. Indeed, at high fields, the characteristic pseudo-period increases with increasing $T$, as expected for conventional UCFs. In this case, the accumulated phase is proportional to the magnetic field flux enclosed by the loops contributing to the interferences. As mentioned above, increasing temperature reduces the loop characteristic size and, in turn, the magnetic field flux (i.e., the accumulated phase) for a given applied field. A larger field variation is therefore needed to accumulate a $2\pi$ phase and complete a quasi-period, causing $B_c$ to increase with temperature. On the contrary, as captured in Fig. \ref{Fig.6}(a), in the fast fluctuations regime $B_c$ is one order of magnitude smaller and is insensitive to the temperature. We argue that this temperature independence is consistent with a geometrical Berry phase. This phase is accumulated by the carriers as they travel around a closed loop in $k$-space enclosing an effective magnetic monopole. The loop size is roughly defined by $k_F$ and is therefore temperature independent. This is in sharp contrast to the dynamical phase accumulated during the transport over a loop in \textit{real} space, whose size is given by the temperature dependent coherence length $L_\Phi$. 

Figure \ref{Fig.6}(b) shows the temperature dependence of $\delta G$ for the high- and low-field regimes. The  variation  is similar for both regimes, with $\delta G$ decreasing with temperature in either case. This behavior is expected in the high-field regime, where only conventional UCF are visible. $\delta G$ depends on the size of the loops in real space producing interference. Again, these loops have a typical length given by $L_\Phi$, which decreases with increasing temperature.\cite{Data_book} This leads to vanishing interference, and hence vanishing oscillations, above 1~K.\cite{Beenakker_review} In the low-field regime, the fast fluctuations also disappear at high temperature, even though these fluctuations are associated with a geometrical phase. This is related to the sensibility of our UCF interferometer.  In our experiments, the Berry phase term is probed through the interference patterns produced by the wave functions of the quasi-particles describing different \textit{real} space paths. The visibility of the interferences is thus given by the coherence length $L_\Phi$. Hence, the fast-fluctuation regime is also lost at high temperature, above 1 K.

\section{Conclusion}

To conclude, we found that GaMnAs is a unique material for addressing Berry phase physics in condensed matter: the combination of ferromagnetism, strong spin-orbit coupling and semi-conducting like nature provides a direct way to control its band structure (hence its Berry phase) by controlling the magnetization of the device. Our measurements give clear indications that the fast-fluctuation regime observed at low magnetic fields is the fingerprint of the change in the band-structure Berry phase, yielding to a direct experimental
evidence of this fundamental mechanism in solid state matter. The next step will be to measure directly the phase itself. Such an experiment will require having good control of the interfering trajectories by using a real interferometer such as an Aharonov-Bohm ring. Using this approach in such a class of materials will provide a direct way to explore one of the most fundamental and elusive concepts of condensed-matter physics.

\subsection*{Acknowledgments}
The authors gratefully acknowledge the LPN clean room support staff.


\begin{thebibliography}{28}
\bibitem{TvsI} The cryostat base temperature was kept constant and the experiments were carried out with different applied currents, ranging between 1 nA and 500 nA and giving rise to different effective temperatures, as discussed in Ref. [16]. The relationship between the electronic temperature and the applied current was accurately checked by comparing the conductance of the sample with calibration experiments at low current.
\end{thebibliography}
\end{document}